\documentclass[journal]{IEEEtran}

\ifCLASSINFOpdf
\else
\fi

\usepackage{graphicx}
\usepackage{epstopdf}
\usepackage{multicol}
\usepackage{url}
\usepackage{amsmath}
\usepackage{amsthm}
\usepackage{mathtools}
\usepackage{amssymb}
\usepackage{esint}
\usepackage{amsfonts}
\usepackage{multirow}
\usepackage{wrapfig}
\usepackage{cite}
\usepackage{subcaption}
\usepackage{lipsum}
\usepackage{float}
\usepackage[none]{hyphenat}
		\usepackage[usenames,dvipsnames]{color}

\begin{document}

\title{IoT Localization for Bistatic Passive UHF RFID Systems with 3D Radiation Pattern\thanks{A conference version of this work is published in IEEE Wireless Communicating and Networking Conference (WCNC) 2015 Proceedings\cite{ciftler2015fundamental}.}}
\author{\IEEEauthorblockN{Bekir Sait \c{C}iftler\IEEEauthorrefmark{1},
Abdullah Kadri\IEEEauthorrefmark{4}, and
\.{I}smail G\"{u}ven\c{c}\IEEEauthorrefmark{7}}
\IEEEauthorblockA{\\\IEEEauthorrefmark{1}Department of Electrical and Computer Engineering, Florida International University, Miami, FL, USA}
\IEEEauthorblockA{\\\IEEEauthorrefmark{7}Department of Electrical and Computer Engineering, North Carolina State University, Raleigh, NC, USA}
\IEEEauthorblockA{\\\IEEEauthorrefmark{4}Qatar Mobility Innovations Center (QMIC), Qatar University, Doha, Qatar}}
\maketitle
\begin{abstract} Passive Radio-Frequency IDentification
(RFID) systems carry critical importance for  Internet of Things (IoT) applications due to their energy harvesting capabilities.
RFID based position estimation, in particular, is expected to facilitate a wide array of location based services for IoT applications with low-power requirements. In this paper, 
considering monostatic and bistatic configurations and 3D antenna radiation pattern, we investigate the accuracy of received signal strength based wireless localization using passive ultra high frequency (UHF) RFID systems.
The Cramer-Rao Lower Bound (CRLB) for the localization accuracy is derived, and is compared with the accuracy of maximum likelihood estimators for various RFID antenna configurations.
Numerical results show that due to RFID tag/antenna sensitivity, and the directional antenna pattern, the localization accuracy can degrade at \emph{blind} locations that remain outside of the RFID reader antennas' main beam patterns.
In such cases optimizing elevation angle of antennas are shown to improve localization coverage, while using bistatic configuration improves localization accuracy significantly.
\end{abstract}


	\section{Introduction}
    Radio Frequency IDentification (RFID) is a promising technology for the proliferation of Internet of Things (IoT) applications, and it can be used to detect and identify the items in the proximity\cite{yan2008internet,welbourne2009Building,atzori2010internet,guvenc2011reliable,Akkaya2015IoT}.
	Due to their cost effective, durable, and energy efficient operation~\cite{access}, RFID technology has been  used in wide range of  applications such as asset management \cite{asset}, access control\cite{acontrol}, public safety\cite{raty2010surveypublic}, localization\cite{localization}, and tracking \cite{tracking}.
	Among these, enabling high accuracy localization  for massively deployed IoT devices carries critical importance for a diverse set of IoT applications~\cite{kortuem2010smart}.

	Localization using radio frequency (RF) signals has been actively researched in the literature over the past decades\cite{Patwari2005July,Gezici2005July,Alimpertis2014Community,guvenc2012fundamental}.
	Outdoor localization is mostly handled with Global Positioning System (GPS) technology whereas indoor localization requires alternative approaches since GPS needs a line-of-sight connection between user equipment and satellites.
  Moreover, massive deployment of IoT devices necessitates energy and cost efficient localization methods for prolonged durations.
	The RFID technology hence becomes a promising alternative for cost-effective, energy efficient  indoor identification and localization for massively deployed IoT.
	
      \begin{figure}[t]
		\centering
\includegraphics[width = 7.8cm]{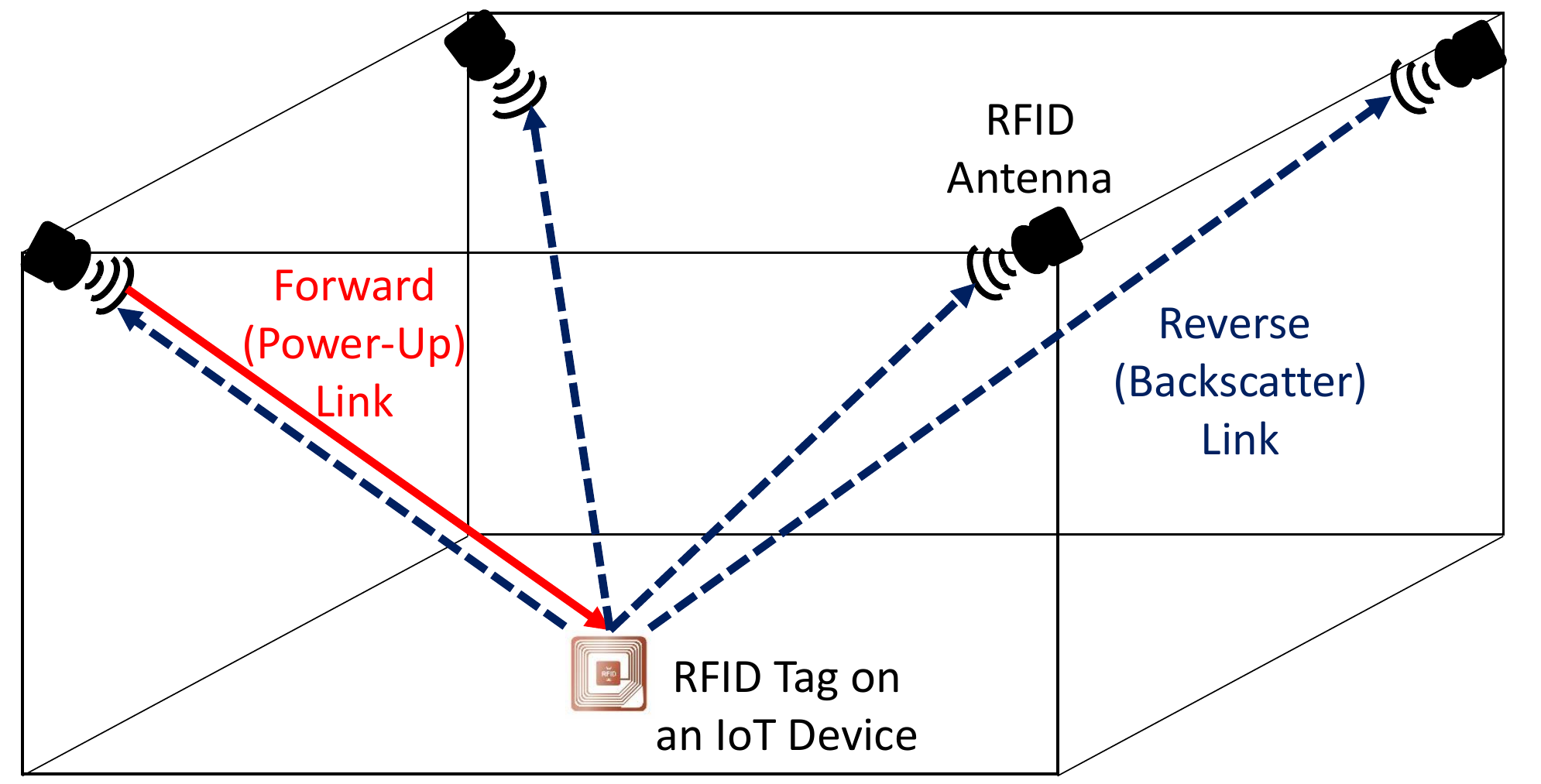}
		\caption{Passive RFID localization system with bistatic configuration. In monostatic configuration, reverse link is available only for the RFID antenna establishes forward link.}\label{fig:IoTFramework}
        \vspace{-7mm}
		\end{figure}  
    
    An Ultra High Frequency (UHF) RFID communication is fundamentally different from the conventional RF communication since it has two distinct links: the forward (power-up) and the reverse (backscatter) link.
	The forward link powers the passive RFID tags and the reverse link carries the information of tags.        
	Ability to power-up tags in the forward link enables \emph{battery-less} operation of RFID tags\cite{battery}, which is a major advantage of RFID systems for low-power IoT applications.
In general, there are two configurations for UHF RFID systems: 1) monostatic configuration, and 2) bistatic configuration. In the monostatic configuration, a single reader antenna transmits the continuous wave, which powers up the passive tag, and subsequently receives the backscattered information signal from the tag. In the bistatic configuration the transmission and reception are handled by different reader antennas as shown in Fig.~\ref{fig:IoTFramework}. These antennas might be co-located (i.e., at same location, closely spaced) or dislocated (at separate locations). A particular challenge with both configuration is that complex, directional, and three dimensional RFID propagation models need to be explicitly taken into account to accurately characterize the real-world forward/backward propagation channels.

	
	In this paper, we use sophisticated and realistic 3D path-loss and radiation models to study fundamental lower bounds on the localization accuracy of Received Signal Strength (RSS) based UHF RFID localization systems for both monostatic and bistatic configurations. The main contributions of this work are as follows: 1)  Cramer-Rao Lower Bound (CRLB) on the localization accuracy are derived in closed-form considering an enhanced RSS model, using the \emph{directional} and \emph{3D} radiation pattern from UHF RFID reader antennas, and the concept of \emph{localization coverage}; 2)  Tag and reader sensitivity is incorporated into analytic derivations both for monostatic and bistatic scenarios, to derive \emph{localizability} and \emph{localization coverage} metrics; 3) Extensive computer simulations are carried out to compare the localization accuracy of the maximum likelihood  (ML) technique 
with the CRLBs, considering directional radiation patterns and using different configurations for RFID reader antennas.

Our analysis and simulation results show that for certain scenarios, using bistatic antenna configuration as in Fig.~\ref{fig:IoTFramework} may increase the average localization coverage by $38\%$ when compared to monostatic RFID configuration.
Another important parameter in the antenna configurations is the elevation angle $\theta$.
Especially with lower transmit powers, it affects the localization coverage and accuracy.
Corner placement of antennas for $\theta=\pi/4$ with $1000$~mW gives $29\%$ localization coverage, while $\theta=\pi/3$ and $\theta=\pi/2$ results in $78\%$.
Our results for the specific RFID configuration show that it is possible to locate a tag within $1$~meter error with a probability of $0.76$ with corner placement of antennas, whereas this probability drastically reduces to $0.53$ when side placement is used for $\theta=~\pi/4$ with bistatic configuration.

	The rest of this paper is organized as follows.
	Literature review for RSS-based localization in passive UHF RFID systems is provided in Section \ref{sect:LitRev}.
	In Section \ref{sect:SystemModel}, the system model is described in detail which involves a 3D radio propagation model for RFID systems.
	The concept of localizability is defined, as well as localization coverage percentage in Section~\ref{sect:CovAreas}.
	Section~\ref{sect:CRLB} derives the CRLBs and the Maximum Likelihood Estimator (MLE) based on the likelihood function for an RFID tag's location for the considered RFID scenario.
	Numerical results are provided in Section~\ref{sect:NResults}, and concluding remarks are given in Section~\ref{sect:Conclusion}.
    
\section{Literature Review}
		\label{sect:LitRev}
        
            	\begin{figure*}[t]
		\centering
		\begin{subfigure}[t]{.35\linewidth}
			\centering			\includegraphics[width=0.8\linewidth]{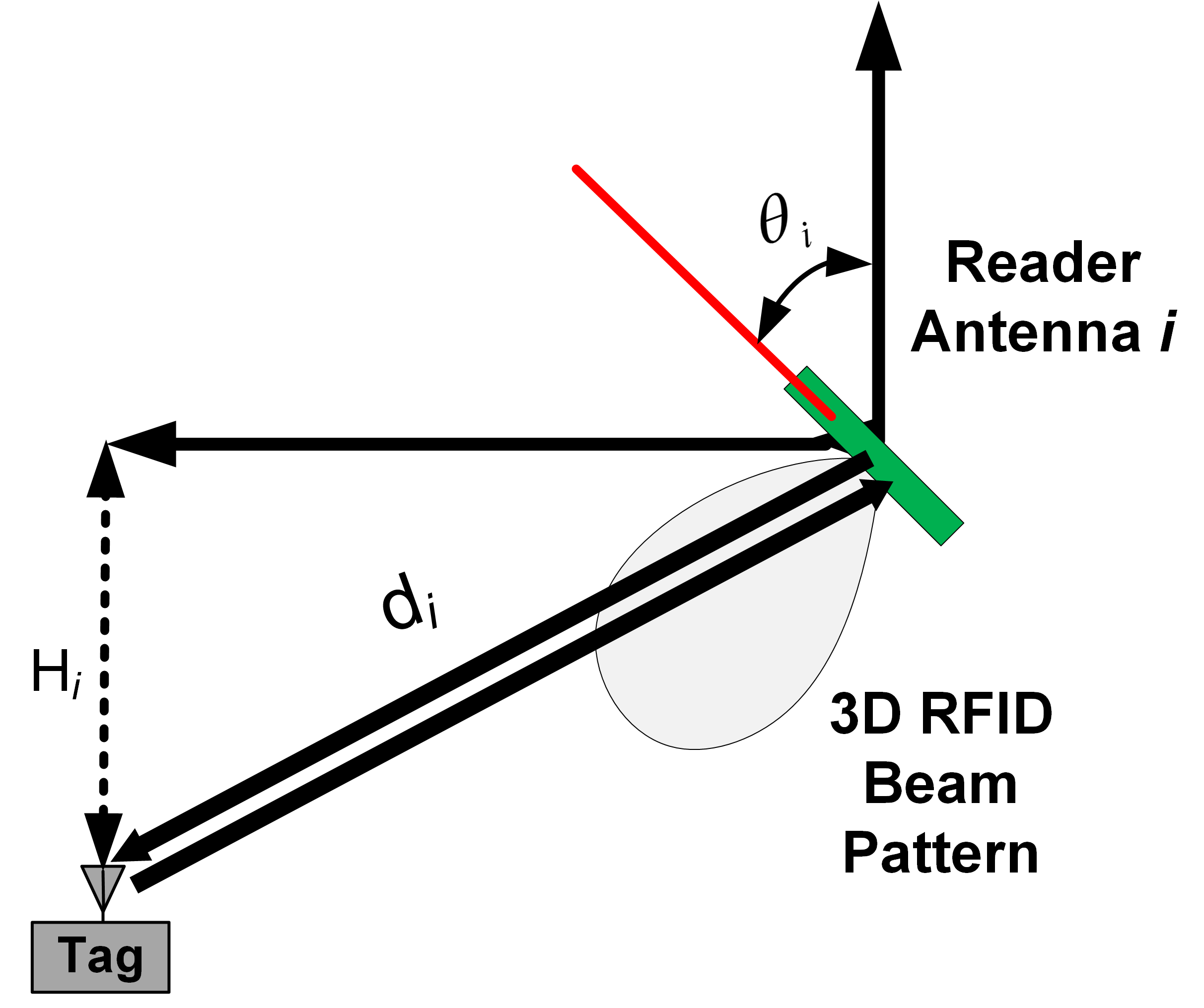}
\caption{}
			\label{fig:SystemModelmono}
		\end{subfigure}~
		\begin{subfigure}[t]{.55\linewidth}
			\centering
			\includegraphics[width=0.8\linewidth]{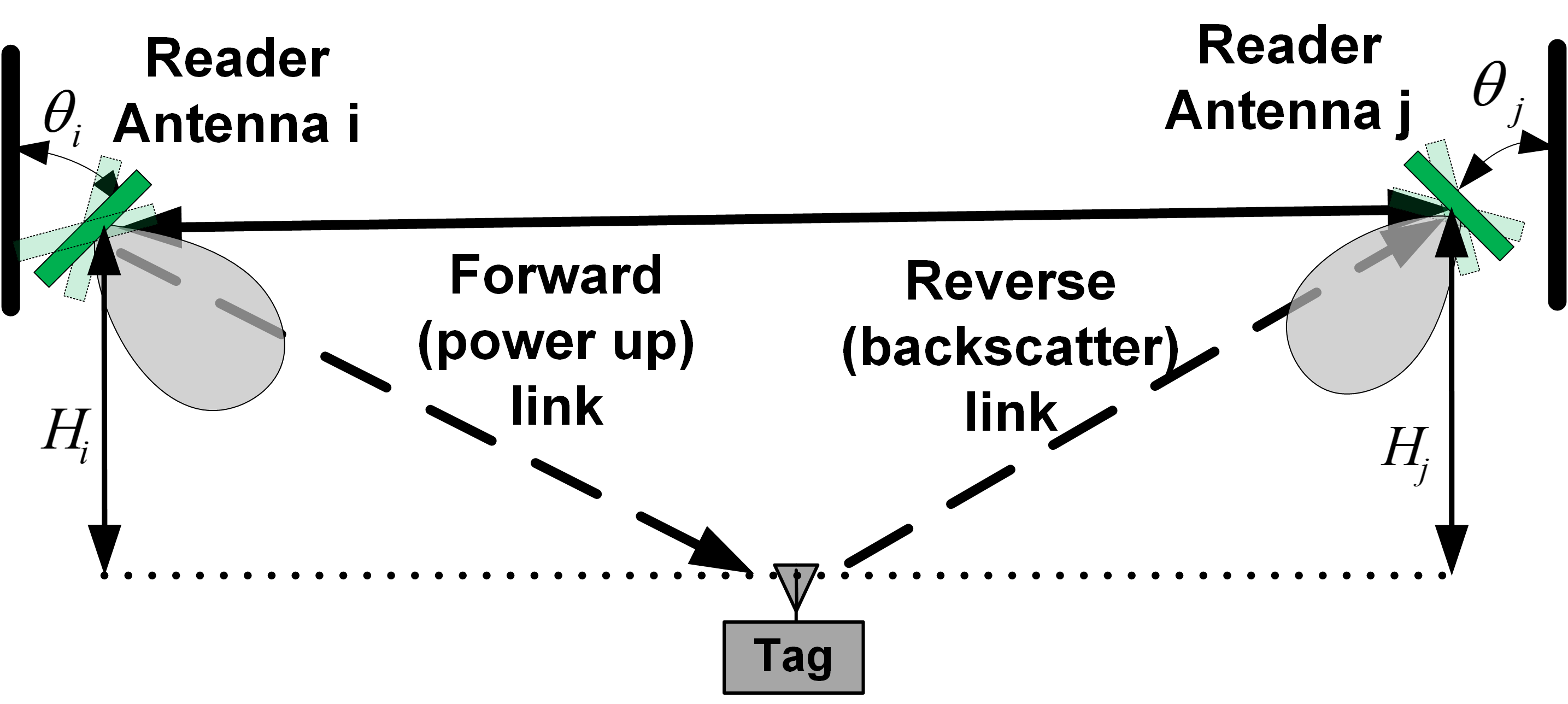}
			\caption{ }
			\label{fig:SystemModel}
		\end{subfigure}
		\caption{(a) In the monostatic configuration, the signal transmitted by reader antenna (tilted by angle $\theta_{i}$) powers the  tag in the forward link. The backscatter link signal, which carries the information of the tag, is received back at the same reader antenna. (b) In the bistatic configuration, the signal transmitted by reader antenna~$i$ (tilted by angle $\theta_{i}$) powers the  tag in the forward link. The backscatter link signal, which caries the information of the tag, is received at reader antenna~$j$ (tilted by angle~$\theta_{j}$).}
			\vspace{-5mm}
		\label{fig:system}
	\end{figure*}    
        
		Although there are several studies in the literature that investigate RSS-based localization with RFID technology \cite{Miesen2011Where,Ni2011RFID,Bouet2008RFID,Li2012Real,6296957}, fundamental lower bounds on RFID-based wireless localization are relatively unexplored.
		In \cite{Hahnel2004Mapping}, authors used a mobile robot with RFID reader antennas to generate map of an indoor environment with RFID tags on the walls.
		After the mapping phase, the robot may locate itself inside the building based on the closest tag information.
		In the LANDMARC localization technique introduced in \cite{Ni2003Landmarc}, reference RFID tags are used for implementing RSS-based indoor localization method, where fixed-location reference tags with known locations are used  to localize the tags.
		In \cite{Zhao2007Vire}, authors improve LANDMARC approach to tackle with multipath effects and RF interference.
		A probabilistic RFID map-based technique with Kalman filtering is used to enhance the location estimation of the RFID tags in \cite{Bekkali2007RFID}.
		Another approach to localize the RFID tags is studied in \cite{Hekimian2010Accurate}, which uses the phase difference information of backscattered signal of the RFID tags.
		In \cite{Leitinger2014Performance}, authors consider a multipath environment to derive the CRLBs on the position error of an RFID based wireless localization system.
		Geometry of the deterministic multipath components and the interfering diffuse multipath components are considered in the backscatter channel model.
		
		Typically a simple path-loss model is used for RFID propagation models in the existing literature \cite{Ni2003Landmarc,ciftler2015experimental,Zheng2014Study}, which employs free-space path-loss signal strength model. 
		These models are not capable of accurately capturing the radiation pattern of RFID reader antennas since they are highly directional.
		There are also several experimental studies in the literature related to RSS-based UHF RFID localization systems.
		In \cite{ciftler2015experimental}, an experimentation with passive UHF RFID system is conducted to investigate the relationship between RSS and distance.
		Recently in \cite{Zheng2014Study}, CRLB of RSS-based localization are derived considering a frequency dependent path-loss propagation model, where the model explicitly depends on the  transmit power level and the transmission frequency.
		Accuracy of several localization techniques are compared to CRLB with given path-loss model via simulations and experiments.
		In \cite{akre2014accurate}, authors used $k$-Nearest Neighbor (kNN) algorithm to estimate the location of the target tag from RSS information.
		An experiment involving four antennas and seventy tags is conducted, which resembles to the simulation scenario in our manuscript. 
		It is shown that power control techniques may significantly improve localization accuracy.
		
		Effects of multipath propagation and signal scattering are considered in \cite{lieckfeldt2009exploiting} for passive UHF RFID localization, using MLE and linear least square techniques. 
		A localization algorithm using the differences of RSS values from various tags under same conditions is also proposed. 
		Its performance, which is shown to outperform the kNN algorithm used in LANDMARC\cite{Ni2003Landmarc}.
		A two-parameter path-loss model for UHF RFID systems is constructed in \cite{hasani2014pathloss}, which shows that the RSS of RFID systems are slightly more stable than WiFi RSS values, and this yields more precise location estimates for RFID RSS-based localization.
In our earlier work\cite{ciftler2015fundamental}, we have studied the bounds on RFID localization for monostatic RFID configuration. In this study, our additional contributions include: 1) use of bistatic antenna configuration and different antenna placement which provides a more generalized framework, 2) use of an enhanced RSS model with lognormal distributed noise which yields different CRLB formulations, 3) incorporation of reader antenna and tag sensitivity into theoretical analysis, 4) study of \emph{localization coverage} for RFID tags, outside of which they can not be localized with a reasonable accuracy, and 5) extensive new simulations to study the effects of various parameters and configurations.
        
		\section{System Model}
		\label{sect:SystemModel}
		
		In the rest of this paper, we consider the RFID localization scenario as shown in Fig.~\ref{fig:system}.
		In particular, Fig.~\ref{fig:system}(a) illustrates a monostatic antenna configuration, 
		where the reader antenna is both the transmitter and the receiver.
		On the other hand, the bistatic antenna configuration is shown in Fig.~\ref{fig:system}(b), where one antenna transmits the power-up signal for RFID tag, and the other antenna receives the backscattered signal from the tag.
		We will consider the more general case of bistatic antenna configuration, and study the monostatic configuration as a special case.
For the considered scenario, let $N$ RFID reader antennas be mounted on the walls, located at a height of $z_i$~meters from the ground for the $i$th antenna.
		As shown in Fig.~\ref{fig:system}(b), RFID reader antennas $i$ and $j$ (which are the forward and reverse antennas, respectively) are tilted by an angle $\theta_i$~and~$\theta_j$, respectively, with reference to the  azimuth plane.
		The goal is to localize an RFID tag, which is located at a distance $H_i$ below a reader antenna $i$.

\subsection{Bistatic RFID Configuration}

The total backscattered received power ${P}_{ij}$ at a bistatic configuration of reader antenna $i$ and antenna $j$, which are located at $(x'_i,y'_i,z'_i)$ and $(x'_j,y'_j,z'_j)$, respectively while the position of the tag is $(x_0,y_0,z_0)$, is given by \cite{abekkali}:
	\begin{align}
	{\widetilde{P}}_{ij}(x_0,y_0,z_0)&=\tau\mu_{\rm T}\rho_{\rm L}P_{\rm Tx}G^2_{\rm T}|G_{\rm R}^iG_{\rm R}^jL(d_i)L(d_j)||h_ih_j\Gamma|^2,\label{eq:PrecLinear}
	\end{align}
	which can be written in logarithmic scale as
	\begin{align}
	{P}_{ij}(x_0,y_0,z_0)&[{\rm dBm}]= 20\log_{10}\big(\tau\mu_{\rm T}\rho_{\rm L}P_{\rm Tx}G^2_{\rm T}|h_ih_j\Gamma|^2\big)\nonumber\\&+20\log_{10}\big(G_{\rm R}^i\big)+20\log_{10}\big(G_{\rm R}^j\big)\nonumber\\& +20\log_{10}\big(L(d_i)\big)+20\log_{10}\big(L(d_j)\big),\label{eq:Preceived2}\end{align}
	$\forall i,j \in \{1,\dots,N\}$ where $\tau$ is a coefficient that quantifies the specific data encoding modulation details that can be calculated using power density distribution of the tag's signal.
		
According to the EPCglobal C1G2 specifications \cite{EPCGen2}, any tag in the interrogation zone of the reader can send back its information by reflecting the incoming continuous wave.
The power transfer efficiency $\mu_{T}~\in~[0,1]$ in~\eqref{eq:Preceived2} quantifies how well the tag is impedance-matched to the antenna.
Polarization loss factor $\rho_{\rm L}$ captures the loss due to the mismatch between the polarization of a transmitter antenna and a receiver antenna.
The effective isotropic radiated power {(EIRP)} of the RFID reader antenna is shown as $P_{\rm Tx}$, while
$G^i_{\rm R}$ and $G_{\rm T}$ are the gain of the RFID reader antenna $i$ and tag antenna, respectively, and $L(d_i)$ is the channel pathloss defined by:
\begin{align}
L(d_i)=\bigg(\frac{\lambda}{4\pi d_i}\bigg)^2,
\end{align}
where $\lambda$ is the wavelength of the signal, and $d_i$ is the distance between the tag and the reader antenna~$i$. 
	The transmit power limit of RFID reader antennas, which is critical for coverage of the reader, is 2 W in effective radiated power (ERP) as stated in EPC Gen2 protocol for UHF RFID systems.
	This makes the EIRP limit for RFID readers 35.15 dBm, which is larger than the highest transmit power of 3~W that was used in our simulations\cite{EPCGen2,Griffin2009Complete}.

The forward-link and backscatter-link channels are represented with $|h_i|^2$ and $|h_j|^2$.
	The parameter $\Gamma$ in \eqref{eq:Preceived2} is the differential reflection coefficient of the tag which is a function of the tag antenna gains $G_{\rm T}$, the \textit{radar cross section} RCS denoted by $\sigma_{\rm RCS}$, and the communication wavelength $\lambda$ as follows\cite{abekkali,Griffin2009Complete,Nikitin2006Theory}:
	\begin{equation}
	|\Gamma|^2=\dfrac{4\pi}{\lambda^2|G_{\rm T}|^2}\sigma_{\rm RCS}~.
	\label{eq:gamma2}
	\end{equation}
	In passive UHF RFID applications, the goal is to maximize RCS, which characterizes the scattered power, while still absorbing sufficient power to operate the chip of the tag.
	In our study, we have utilized statistical models for RCS and $\Gamma$ which we obtained from \cite{Nikitin2006Theory,Griffin2009Complete,bletsas2010improving,hasani2014pathloss}.

Assuming a scenario as illustrated in Fig.~\ref{fig:system}, and adopting the expression provided by \cite{abekkali}, a modified \emph{directional} gain of a patch antenna for a 3D propagation environment can be expressed as follows:
\begin{align}
G^i_{\rm R}(\alpha_i,\phi_i)&=3.136\Big[\tan(\alpha_i)\sin\big(0.5\pi\cos(\alpha_i)\big)	\nonumber\\&\times\cos\big(0.5\pi\sin(\alpha_i)\sin(\phi_i)\big)\Big]^2~,
\label{eq:patchantennagain2}
\end{align}
where $\alpha_i=\theta_i-\arcsin(\frac{H_i}{d_i})$, with $\theta_i$ and $\phi_i$ being the elevation and azimuthal angles of the patch antenna $i$, respectively. The parameter $H_i$ in~\eqref{eq:patchantennagain2} is the difference between height of the reader antenna and the height of tag.

\subsection{Translation to Cartesian Coordinate System}

Location of a tag with respect to reader antenna is defined with relative elevation and azimuthal angles, and distance between tag and reader antenna.
On the other hand, derivation of CRLB requires translation from polar coordinate system to the Cartesian coordinate system.
Gain of patch antenna is defined in \eqref{eq:patchantennagain2}, which can be represented in the Cartesian coordinate system as
\begin{align}
&G^i_{\rm R}(x_i,y_i,z_i,x_0,y_0,z_0,\theta_i,\phi_i)\nonumber\\
&=3.136\times\bigg(\frac{\tan\theta_i-\frac{z_i-z_0}{d_i}}{1-\frac{z_i-z_0}{d_i}\tan\theta_i}\bigg)^2\nonumber\\
&\times\sin^2\bigg(\frac{\pi}{2}\bigg(\frac{l_i}{d_i}\cos\theta_i+\frac{z_i-z_0}{d_i}\sin\theta_i\bigg)\bigg)\nonumber\\
&\times\cos^2\bigg(\frac{\pi}{2}\bigg(\frac{l_i}{d_i}\sin\theta_i-\frac{z_i-z_0}{d_i}\cos\theta_i\bigg)\nonumber\\
&\times\bigg(\frac{(x_i-x_0)\cos\phi_i+(y_i-y_0)\sin\phi_i}{l_i}\bigg)\bigg),\label{eq:patchgaincartesian}
\end{align}
where $(x_i,y_i,z_i)$ is coordinate of the antenna-$i$, and $(x_0,y_0,z_0)$ is the location of tag.
The distance between tag and antenna-$i$ is defined with $d_i~=~\sqrt{(x_0-x_i)^2+(y_0-y_i)^2+(z_0-z_i)^2}$, while its projection on the $xy$-plane is given with $l_i~=~\sqrt{(x_0-x_i)^2+(y_0-y_i)^2}$.

\subsection{Monostatic RFID Configuration}
As in Fig.~\ref{fig:system}(a), monostatic RFID is a special case of bistatic RFID configuration, there the transmitter and receiver antenna are identical.
This makes $G_R^i$ and $G_R^j$ equal ($i=j$).
Therefore, the received power in dBm at the reader antenna with monostatic configuration simplifies to:{
	\begin{align}
	{P}_{ii}(x_0,y_0,z_0)&= 20\log_{10}\big(\tau\mu_{\rm T}\rho_{\rm L}P_{\rm Tx}G^2_{\rm T}|h_i|^4|\Gamma|^2\big)\nonumber\\&+40\log_{10}\big(G_{\rm R}^i\big)+40\log_{10}\big(L(d_i)\big),\label{eq:Preceived2mono}
	\end{align}}for $i=1,\cdots,N$.
Note that, for monostatic configuration, ${P}_{ ij}=0$ when $i\neq j$, for all 	$i,j\in \{1,\dots,N\}$.
Therefore, monostatic configuration essentially uses a subset of the antenna reader pairs in bistatic configuration during localization.
In the rest of this paper, bistatic configuration as defined in \eqref{eq:Preceived2} will be assumed to capture measurements at all pairwise combinations of antenna readers, including those corresponding to monostatic configurations.

\section{Tag / Reader Antenna Sensitivity and Localization Coverage}
\label{sect:CovAreas}

\begin{figure*}[t]
	\begin{subfigure}[t]{0.33\textwidth}
		\centering
		\includegraphics[width=.9\linewidth]{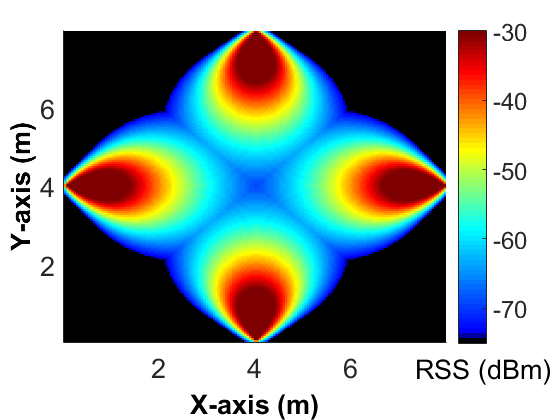}
        \caption{}
		\label{fig:maxRSS45}
	\end{subfigure}%
	~
	\begin{subfigure}[t]{0.33\textwidth}
		\centering
		\includegraphics[width=.9\linewidth]{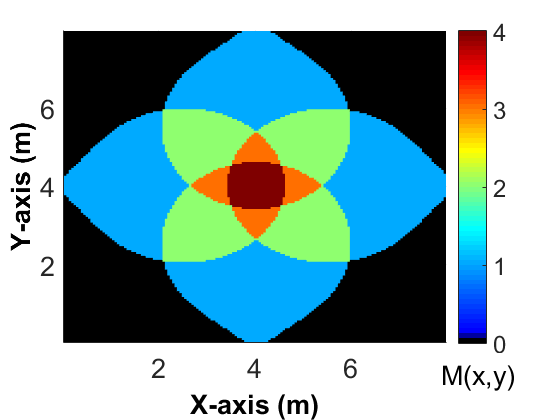}
        \caption{}
		\label{fig:covmono}
	\end{subfigure}
	~
	\begin{subfigure}[t]{0.33\textwidth}
		\centering
		\includegraphics[width=.9\linewidth]{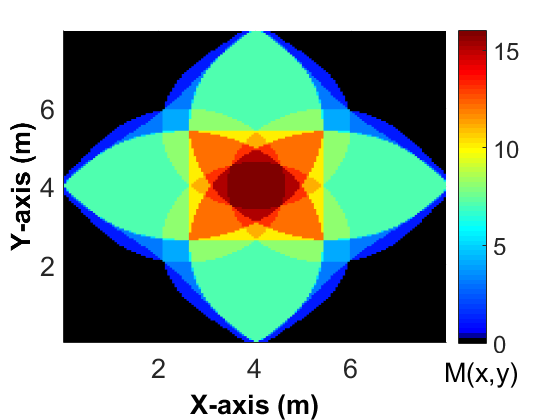}
        \caption{}
		\label{fig:covbi}
	\end{subfigure}
	\caption{(a) Maximum achievable RSS at any possible tag location for the system. (b) Localization coverage for monostatic configuration.   (c) Localization coverage for bistatic configuration. Areas where $M\geq2$ are considered to be \emph{localizable}, and the deployment parameters are $\theta_i=\pi/4$, $H_i=1$ meter for $i=1,2,3,4$, and $P_{\rm Tx}=1000$~mW.}
	\vspace{-5mm}
	\label{fig:t45}
\end{figure*}

In this section, we will first introduce the concepts of \emph{tag antenna sensitivity}, \emph{reader antenna sensitivity}, and \emph{localization coverage} of an RFID system, which corresponds to the spatial region in which an RFID tag will be considered \emph{localizable}. 

\subsection{Tag Antenna Sensitivity} 
\label{subsect:TagAntenna}
The passive tags do not have an internal power structure to modulate or transmit any signal.
They use the received power to both modulate the signal with the internal chip, and backscatter modulated signal to reader.
As one can expect, RFID tags have certain power requirements.
State-of-art tags are able to modulate signals with RSS as low as $-20$ dBm\cite{nikitin2012uhf}, and will not be able to detect the received signal at lower power levels.
The RSS from $i$th reader at tag is defined with
	\begin{align}
	{P}_{i}(x_0,y_0,z_0)&= 20\log_{10}\big(\rho_{\rm L}P_{\rm Tx}G_{\rm T}G^i_{\rm R}L(d_i)|h_i|^2\big)\label{eq:tagSens}
	\end{align}for $i=1,\cdots,N$.
    Note that \eqref{eq:tagSens} is a subset of \eqref{eq:Preceived2} and \eqref{eq:Preceived2mono}  which characterize the RSS after round-trip signal propagation, since \eqref{eq:tagSens} represents only the forward-link.

\subsection{Reader Antenna Sensitivity}
\label{subsect:ReaderAntenna}
In either monostatic or bistatic configuration, an RFID reader must correctly detect the backscattered modulation from the tag, which relies on the reader antenna sensitivity.
Therefore, the received power as in \ref{eq:Preceived2} and \ref{eq:Preceived2mono} must be larger than the reader antenna sensitivity in order to be detected.
The \emph{detection coverage} of an RFID configuration is defined as detectability of a tag at a certain location with that configuration.
The detectability is assumed deterministic with respect to RSS and sensitivity of RFID reader antenna.

\subsection{Coverage Areas for Localization}
\label{subsect:CoverageAreas}
In this subsection, we investigate the impact of the sensitivity of the tag and reader antennas on localization performance.  We introduce below several new metrics for characterizing tag/reader sensitivities and localization coverage.

\textbf{\textit{Definition 1:}} The \textit{coverage} for a given antenna pair $(i,j)$ at a given location $(x,y)$ is captured by a binary deterministic parameter $C_{ ij}(x,y)$, which is defined as:
\begin{equation}
\mbox{$
	C_{{ij}}(x,y)=
	\begin{cases}
	1,& \text{if } {P}_{{ij}}\geq -75 \mbox{ dBm} \mbox{ and } {P}_{{i}}\geq -20 \mbox{ dBm}\\
	0,              & \text{otherwise}
	\end{cases}~.$
}\label{eq:C}
\end{equation}
using \eqref{eq:Preceived2} and \eqref{eq:tagSens}.

Due to nonlinearity of antenna propagation model, in order to localize a tag, at least two different RSS measurements from that tag at a particular position $(x,y)$ are needed.
On the other hand, there might be some tags which are detected from only a monostatic antenna or a bistatic antenna pair, and those tags cannot be localized due to limited information.
Note that for monostatic configuration, $C_{ij}(x,y)=0$ if $i\neq j$. Then  we can define the localizability of a tag as follows.

\textbf{\textit{Definition 2:}} A tag at a given location $(x,y)$ is \emph{localizable} if the following condition is satisfied
\begin{align}
M(x,y)=\sum_{i=1}^{N}\sum_{j=i}^{N}C_{ ij}(x,y)\geq 2~,
\end{align}
where $N$ is the number of antennas in the system, and $M(x,y)$ is the total number antenna configurations that can detect the tag at location $(x,y)$.

\textbf{\textit{Definition 3:}} The \emph{localization coverage} of a tag at position $(x,y)$ is defined with $L(x,y)$ as follows:
\begin{align}
\mbox{$
	L(x,y)=
	\begin{cases}
	1,& \text{if } M(x,y)\geq 2\\
	0,              & \text{otherwise}
	\end{cases}~.$
}\label{eq:L}
\end{align}
The RFID tag can be localized at position $(x,y)$ when $L(x,y)=~1$, and is not localizable when $L(x,y)=0$.

\textbf{\textit{Definition 4:}} The \emph{localization coverage percentage} at a physical area $\mathcal{A}$ can be formally expressed as follows:
\begin{align}
L_p(\mathcal{A})=\dfrac{\iint\limits_{x,y\in\mathcal{A}} L(x,y){\mathrm d}x{\mathrm d}y}{\iint\limits_{x,y\in\mathcal{A}}{\mathrm d}x{\mathrm d}y}\times 100\%~.\label{eq:area}
\end{align}
Note that \eqref{eq:area} defines the percentage of localizable area to total~area. 

In Fig.~\ref{fig:t45}, results from an example deployed scenario for parameters $\theta_i=\pi/4$, $H_i=1$ meter for $i=1,2,3,4$, and $P_{\rm Tx}=1000$~mW are shown. Maximum achievable RSS at any possible tag location is represented in Fig.~\ref{fig:t45}(a), while monostatic and bistatic localization coverage are shown in Fig.~\ref{fig:t45}(b)  and Fig.~\ref{fig:t45}(c), respectively.
The localization coverage percentage is $21\%$ for monostatic configuration, while it is above $50\%$ for bistatic configuration.
The number of maximum measurements increases from 4 for monostatic configuration to 16 for bistatic configuration.

\section{Cramer-Rao Lower Bound and MLE}
\label{sect:CRLB}

The CRLB is a bound on the variance of any unbiased estimator for an unknown variable, such as the location of an RFID tag, based on a set of observations.  
In this section, we define likelihood function, and derive the CRLB on the accuracy of RSS-based UHF RFID localization systems as a function of various parameters of interest.
We consider both monostatic and bistatic cases for CRLB analysis.
Subsequently, the MLE for unknown RFID tag location is also defined.
Comparison of CRLB and MLE for monostatic and bistatic configurations in various scenarios will be presented through numerical results in Section~\ref{sect:NResults}.

\subsection{Likelihood Function for Unknown RFID Tag Localization}
\label{ssect:likelihood}
When a tag is localizable, then its exact location can be estimated using the measurements obtained at different antenna pairs. 
The probability of an RFID tag being at a certain location can be characterized by its \emph{likelihood function}\cite{Estimation1993Kay}.
Let~$\boldsymbol{\rm x}=~[x,y]$ denote the unknown location of the tag, assuming that the received power in log scale at an RFID reader antenna is subject to Gaussian noise\cite{Patwari2005July}.
Consider that the observations of received power in \eqref{eq:Preceived2} from different RFID antennas mounted on the walls are stacked in a vector $\boldsymbol{\hat{\rm p}}[\mbox{dBm}]$. 
Then, this vector can be modeled as follows
\begin{equation}
\boldsymbol{\hat{\rm p}}=\boldsymbol{\rm p}\boldsymbol{+\omega},~\boldsymbol{\omega} = [\omega_{11},...,\omega_{ij},...,\omega_{NN}]^T,\omega_{ij}\sim\mathcal{N}(0,\sigma^2),
\label{eq:measured}
\end{equation}
where $i=1,...,N$, $j=1,...,N$, and  $\boldsymbol{{\rm p}}$ is a vector of true RSS values which has a size of $N^2$ for a bistatic configuration, and a size of $N$ for a monostatic configuration.
The additive noise on received power, which is assumed independent and identically distributed (iid),  is captured by $\omega_{ij}$, corresponding to the measurement at antenna couple $i$ and $j$, with $\mathcal{N}(\mu,\sigma^2)$ denoting the Gaussian distribution with mean $\mu$ and variance $\sigma^2$.
Then, for the general case of bistatic antenna configuration, the respective likelihood function for the received power at a location $\boldsymbol{\rm x}$ can be written as:
\begin{align}
\mathcal{L}(\boldsymbol{\rm \hat{p}};\mbox{$\boldsymbol{\rm x}$})&=\dfrac{1}{(2\pi\sigma^2)^{\frac{N^2}{2}}}\nonumber\\
&\times\exp\left({-\frac{1}{2\sigma^2}\sum_{i=1}^{N}\sum_{j=1}^{N}C_{ij}(x,y)\left({P}_{ ij}-\hat{{P}}_{ ij}\right)^2}\right), \label{eq:likelihood}
\end{align}
where ${P}_{ij}$ is the value of RSS for reader antennas $i$ and $j$, and it depends on the unknown tag location $\boldsymbol{\rm x}=(x,y)$ as defined in \eqref{eq:Preceived2}. For the monostatic configuration, it can be easily shown that \eqref{eq:likelihood} simplifies to the following likelihood function:
\begin{align}
\mathcal{L}(\boldsymbol{\rm \hat{p}};\mbox{$\boldsymbol{\rm x}$})=\dfrac{1}{(2\pi\sigma^2)^{\frac{N}{2}}}
\exp\left({-\frac{1}{2\sigma^2}\sum_{i=1}^{N}C_{ii}(x,y)\left({P}_{ ii}-\hat{{P}}_{ ii}\right)^2}\right). \label{eq:likelihoodm}
\end{align}
\vspace{-5mm}

\subsection{CRLB Analysis}
\label{ssect:CRLBanalysis}
Based on the 3D and directional propagation model defined in \eqref{eq:PrecLinear}--\eqref{eq:patchgaincartesian},  the localization coverage parameter $C_{ij}(x,y)$ defined in \eqref{eq:C}, and the likelihood function defined in \eqref{eq:likelihood} the CRLB on the variance of an unbiased estimator for $\boldsymbol{\rm x}$ can be defined as follows.

\newtheorem{prop}{Theorem}\label{prop:p1}
\begin{prop}
	The CRLB on the root mean square error (RMSE) of an unbiased position estimator $\boldsymbol{\hat{\rm x}}$ based on the measurements model in \eqref{eq:measured}  and the likelihood function in \eqref{eq:likelihood} is given by:\vspace{-5mm}
	\end{prop}
	\begin{align}
	\mbox{RMSE}_{\mbox{loc}}(x,y)\geq\sqrt{\textbf{I}^{-1}_{11}+\textbf{I}^{-1}_{22}}~,
	\label{eq:CRLB}
	\end{align}
	where $[\textbf{I}(\mbox{$\boldsymbol{\rm x}$})]$ is the Fisher Information Matrix (FIM) for $\mbox{$\boldsymbol{\rm x}$}$,
	\begin{equation}
	[\textbf{I}(\mbox{$\boldsymbol{\rm x}$})] = \begin{bmatrix} \textbf{I}_{11} & \textbf{I}_{12} \\ \textbf{I}_{21} & \textbf{I}_{22} \end{bmatrix},
	\label{eq:fim}
	\end{equation}
	whose elements are as derived in~\eqref{eq:fim_element}-\eqref{eq:partialdx}.
	
	\textit{Proof:} See Appendix~\ref{sect:Derivation}.~\hfill$\qed$
	
	An example derivation of the CRLB for the special case of $\theta=\pi/4$ for all $i$ is explained in detail in Appendix~\ref{sect:DerivationExample}.

\subsection{Maximum Likelihood Estimator}
While the CRLB gives a lower bound on the localization RMSE, an effective estimator is needed to find an RFID tag's location as accurate as possible, ideally with an RMSE close to the CRLB. 
In here, we will define a simple MLE estimator for comparison purposes with the CRLB.
Using the likelihood function defined in \eqref{eq:likelihood}, the MLE can be formulated as follows~\cite{Estimation1993Kay}
\begin{align}
\hat{\boldsymbol{\rm x}} = \arg \max_{\mbox{$\boldsymbol{\rm x}$}}~~ \mathcal{L}(\boldsymbol{\rm P};\mbox{$\boldsymbol{\rm \hat{x}}$})~.
\label{eq:mle}
\end{align}
Having a closed form solution for the MLE in \eqref{eq:mle} is not mathematically tractable due to the complexity of the directional antenna radiation pattern as captured through \eqref{eq:Preceived2}-\eqref{eq:patchgaincartesian}. 
In particular, due to entangled sines and cosines, after equating differentiation of the likelihood function as in \eqref{eq:partialdx} to zero, one cannot obtain a closed form solution.
Thus our problem could be solved with MLE grid search, which can be represented as follows
\begin{align}
\hat{\boldsymbol{\rm x}} = \arg \min_{\mbox{$\boldsymbol{\rm x}$}}~~ \sum_{i=1}^{N}\sum_{j=1}^{N}\bigg({P}_{ ij}(x,y)-\hat{{P}}_{ ij}(x,y)\bigg)^2.
\label{eq:mlesearch}
\end{align}


In our computer simulations in Section~\ref{sect:NResults}, we consider a densely sampled grid of nearly $15000$ uniformly spaced points. 
The granularity of the grid is set to $5$ cm. 
	Then, the MLE solution corresponds to the grid position that maximizes the likelihood function~in~\eqref{eq:likelihood} and can be found using exhaustive search. 
	To reduce complexity, the~MLE solution is found by a constrained search over the region that is defined by the number of RSS measurements and corresponding antennas.
	When there are only two RSS measurements available, the search is conducted only over the positions where $M(x,y)~=~2$. 
	As it is stated in Section \ref{subsect:CoverageAreas}, a grid location with only two RSS measurements is still localizable, although the accuracy is relatively limited when compared to locations where more than two RSS measurements are available.
	Based on our numerical results that will be shown in Section~\ref{sect:NResults}, overall localization accuracy is still acceptable. Accuracy of the MLE will be compared with the CRLB in various scenarios in the next section.

\begin{table}[t]
	\centering
	\caption{Passive UHF RFID system parameters.}
	\begin{tabular}{cc}
		\hline
		\multicolumn{1}{|l|}{\textbf{Parameter}} & \multicolumn{1}{l|}{\textbf{Value}} \\ \hline\hline
		\multicolumn{1}{|l|}{Operating Frequency} & \multicolumn{1}{l|}{$865.7$ MHz} \\ \hline
		\multicolumn{1}{|l|}{Operating Bandwidth} & \multicolumn{1}{l|}{$300$ kHz} \\ \hline
		\multicolumn{1}{|l|}{Transmit Power ($P_{\rm Tx}$) {(EIRP)}} & \multicolumn{1}{l|}{$1000$~mW to $3000$~mW} \\ \hline
		\multicolumn{1}{|l|}{Modulation Efficiency ($\tau$)} & \multicolumn{1}{l|}{$0.5$} \\ \hline
		\multicolumn{1}{|l|}{Polarization Loss Factor ($\rho_{\rm L}$)} & \multicolumn{1}{l|}{$0.5$} \\ \hline
		\multicolumn{1}{|l|}{Power Transfer Efficiency ($\mu_{\rm T}$)} & \multicolumn{1}{l|}{{$0$ to $1$}} \\ \hline
		\multicolumn{1}{|l|}{Differential Reflection Coefficient ($|\Gamma|^2 $)} & \multicolumn{1}{l|}{{$0$ to $1$}} \\ \hline
		\multicolumn{1}{|l|}{Tag Antenna Gain ($G_{\rm T} $)} & \multicolumn{1}{l|}{$0$ dBi} \\ \hline
		\multicolumn{1}{|l|}{Tag Antenna Sensitivity ($G_{\rm T} $)} & \multicolumn{1}{l|}{$-18$ dBm} \\ \hline
		\multicolumn{1}{|l|}{Reader Antenna Sensitivity ($R_{\rm S} $)} & \multicolumn{1}{l|}{$-75$ dBm} \\ \hline
		\multicolumn{1}{|l|}{Antenna Height} & \multicolumn{1}{l|}{$2$ m} \\ \hline
		\multicolumn{1}{|l|}{Tag Height} & \multicolumn{1}{l|}{$1$ m } \\ \hline
		\multicolumn{1}{|l|}{Room Width and Length} & \multicolumn{1}{l|}{$8$ m x $8$ m} \\ \hline
		\multicolumn{1}{|l|}{{Granularity of Simulations}} & \multicolumn{1}{l|}{{$1$ cm}} \\ \hline
		\multicolumn{1}{|l|}{Reader Antenna Elevation Angle ($\rm\theta $)} & \multicolumn{1}{l|}{$\pi$/4 to $\pi$/2 } \\ \hline
		\label{table:parameter}
	\end{tabular}
	\vspace{-5mm}
\end{table}


\section{Numerical Results}
\label{sect:NResults}

Numerical results are provided to validate analytic derivations with computer simulations and to compare the performance of the MLE with the CRLB for RFID based IoT localization.
The simulation parameters for the passive UHF RFID system is given in Table~\ref{table:parameter}.
As stated in Section \ref{sect:CRLB}, the received power at the RFID reader antenna is subject to lognormal noise.
	The noise variance is adopted from~the~statistical models in~\cite{hasani2014pathloss,chawla2013localization}, which were derived from RFID propagation measurements.  
    
Our computer simulation considers RFID antennas that are installed in a square shaped room with $8$ meters width, and the height of the reader antennas are $2$ meters above floor level.
The channel is assumed to be frequency flat slow fading channel in our system.
There are two antenna placement configurations, one is placing the antennas to centers of side walls which is referred as `Side', and the second is placing them on the corners of the room which is referred as `Corner' in figures.
The reader uses circularly polarized antennas which have a radiation pattern as defined in \eqref{eq:patchantennagain2}, and the tag antennas are assumed to be vertically polarized.
The height of the tag is assumed to be known and $1$ meter.
Elevation angles of reader antennas are defined as $\pi/4$, $\pi/3$, and $\pi/2$.
Elevation angles lower than $\pi/4$ are not considered due to lack of localization coverage for those angles.

\subsection{Localization Coverage}

\begin{figure*}[!t]
	\centering
	\includegraphics[width=0.85\linewidth]{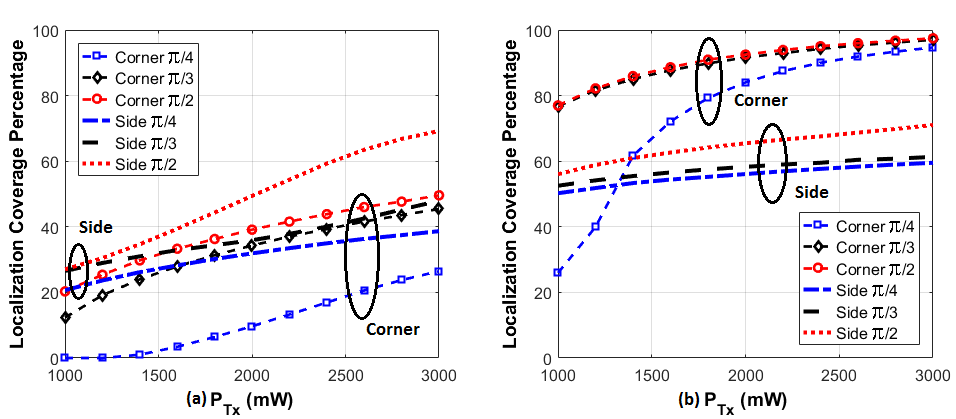}
	\caption{Localization coverage percentage for $\theta_i=\pi/4$, $\pi/3$, $\pi/2$, and $i=1,2,3,4$, (a) monostatic and (b) bistatic configurations.}
	\label{fig:CoverageP}
\end{figure*}

In Fig.~\ref{fig:CoverageP}, localization coverage percentage in \eqref{eq:area} is illustrated for different elevation angles, antenna placement configurations, and transmit power levels for monostatic (Fig.~\ref{fig:CoverageP}(a)) and bistatic (Fig.~\ref{fig:CoverageP}(b)) antenna configuration.
The localization coverage is below $50\%$ for monostatic cases other than Side~$\pi/2$.
The coverage percentage for monostatic configuration increases rapidly with increasing transmit power from $17.8\%$ on the average for $P_{\rm Tx}~=~1000$~mW to $46.2\%$ for $P_{\rm Tx}~=~3000$~mW transmit power.
Localization coverage for bistatic cases show improvement with increased transmit power as well.
The mean localization coverage percentage for $P_{\rm Tx}~=~1000$~mW is $56.4\%$, while increasing transmit power to $P_{\rm Tx}~=~3000$~mW substantially boosts it to $80.2\%$.

The elevation angle also plays a critical role in localization coverage of the system.
In monostatic and bistatic configurations, $\theta=\pi/2$ is superior to other angles for both Corner and Side placement of antennas.
In general, the coverage is increased with increased elevation angle.
Corner placement of the antennas is better in bistatic configuration, whereas in monostatic configuration side placement has larger coverage area in general.
The corner placement of the antennas covers $85\%$ of the area for bistatic configuration on the average for all available transmit powers, whereas side placement enables to localize the tags in $59.4\%$ of the area.
Things are different for monostatic case, where corner placement has $27\%$ coverage, while side placement achieves better performance with $38.9\%$.
This is expected since side placement increases the overlap possibility of monostatic antenna coverages with less distances between antennas, whereas corner placement exploits the radiation coverage with increased distances between antennas.

\begin{figure}[!t]
	\centering
	\begin{subfigure}{0.49\linewidth}
		\centering
		\includegraphics[width=.99\linewidth]{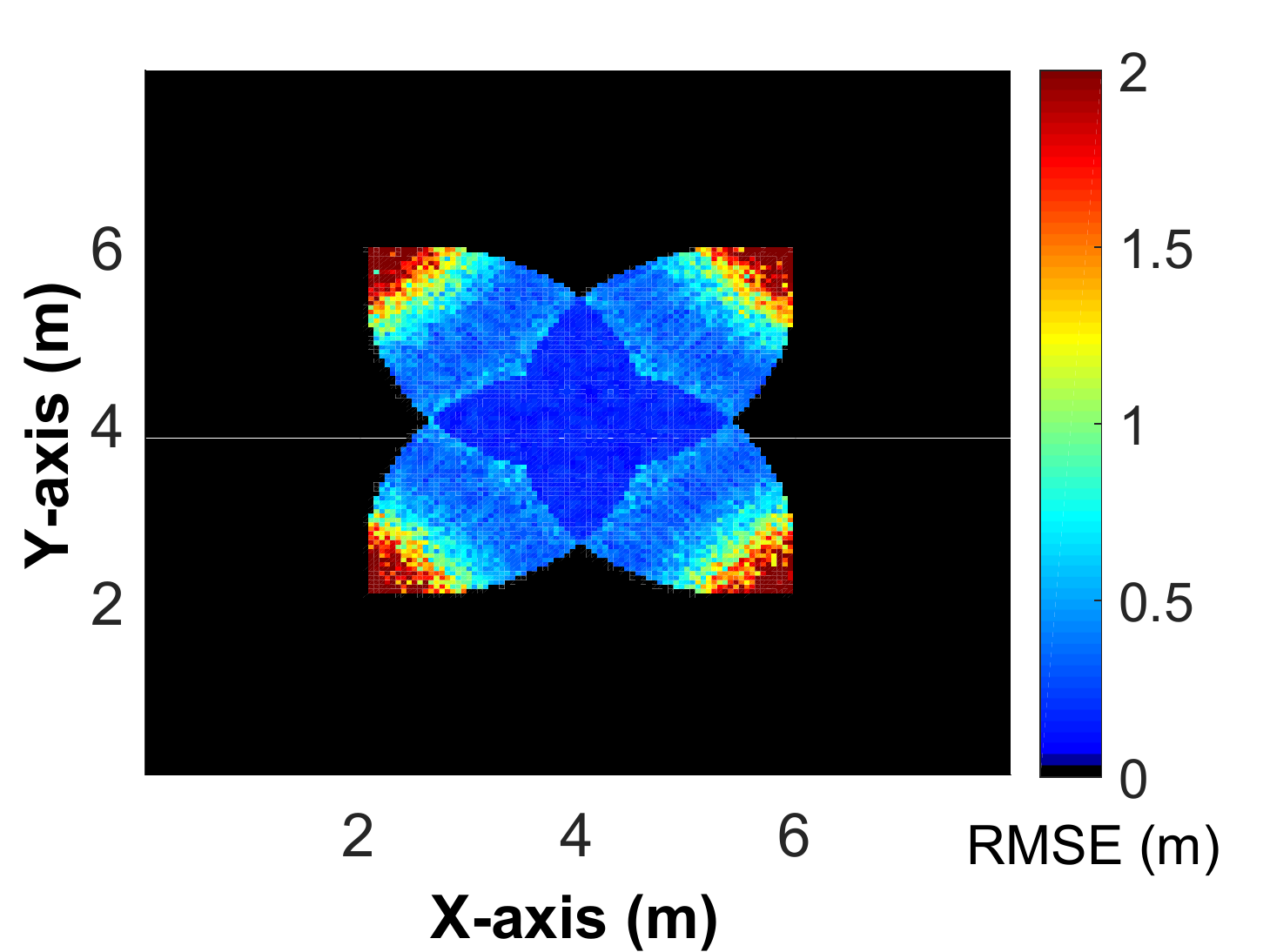}
		\vspace{-5mm}
		\label{fig:mle45m}
		\caption{MLE Monostatic}
	\end{subfigure}~\begin{subfigure}{0.49\linewidth}
	\centering
	\includegraphics[width=.99\linewidth]{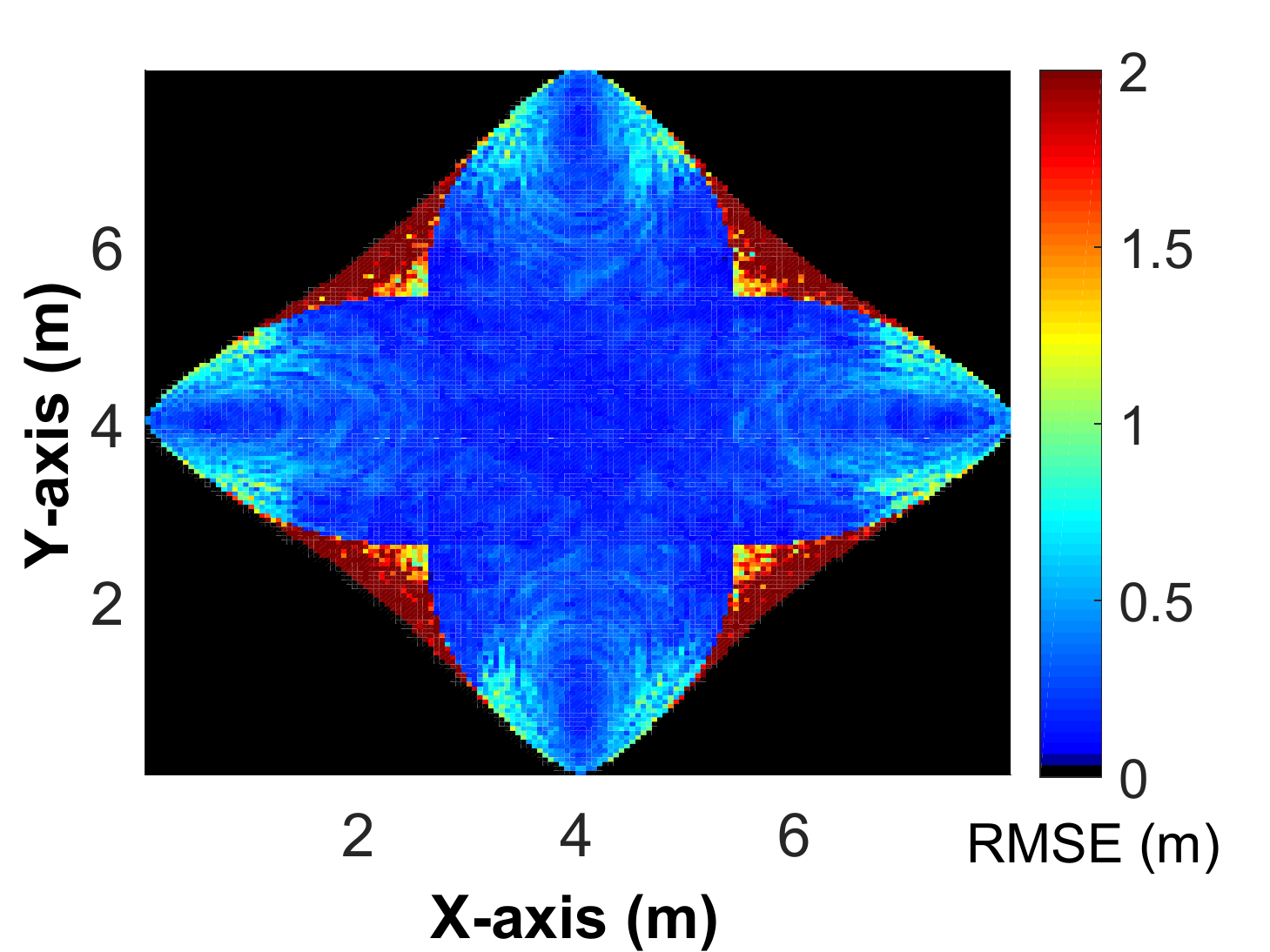}
	\vspace{-5mm}
	\label{fig:mle45b}
	\caption{MLE Bistatic}
\end{subfigure}
\\
\begin{subfigure}{0.49\linewidth}
	\centering
	\includegraphics[width=.99\linewidth]{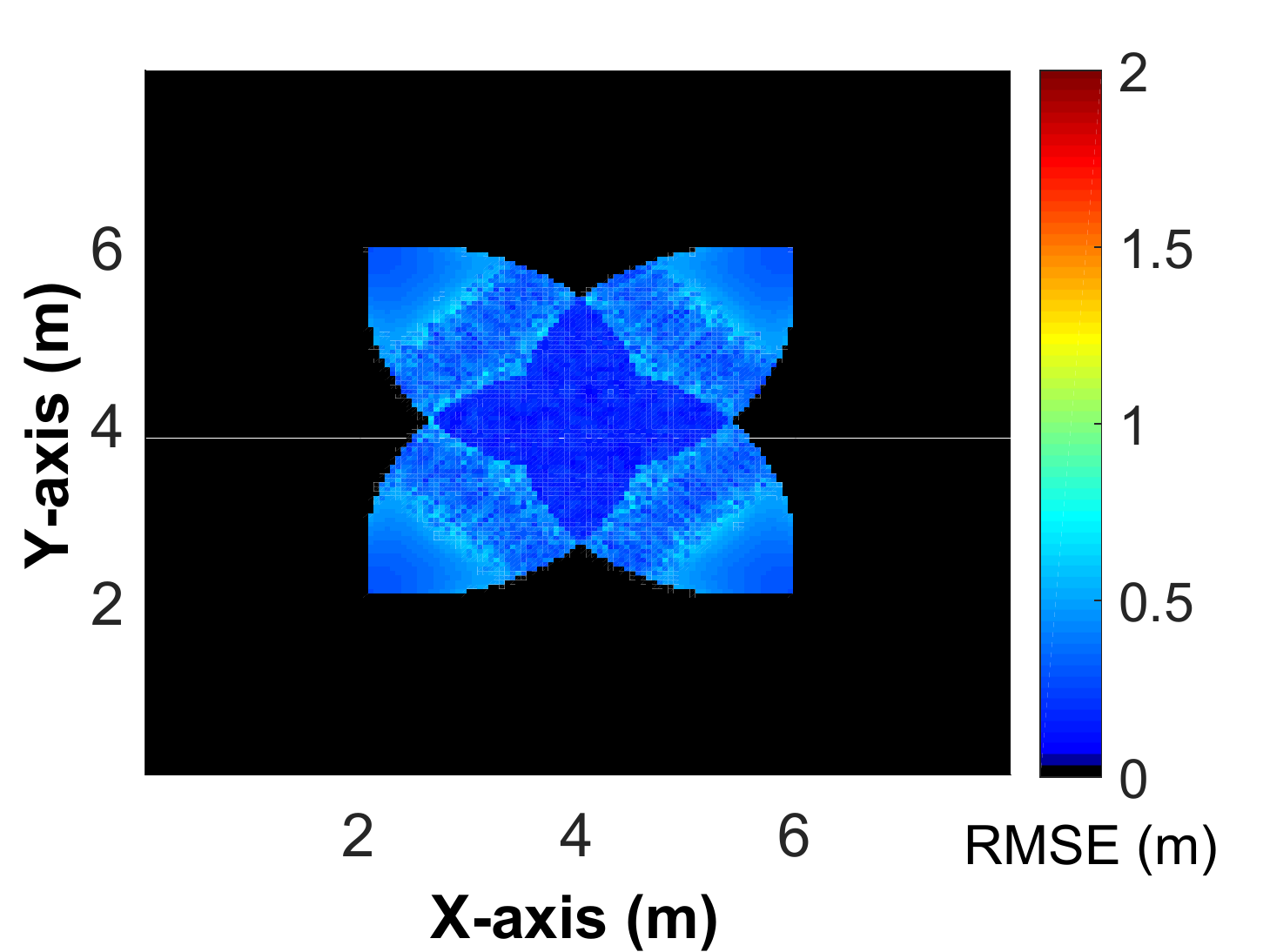}
	\vspace{-5mm}
	\label{fig:crlb45m}
	\caption{CRLB Monostatic}
\end{subfigure}~\begin{subfigure}{0.49\linewidth}
\centering
\includegraphics[width=.99\linewidth]{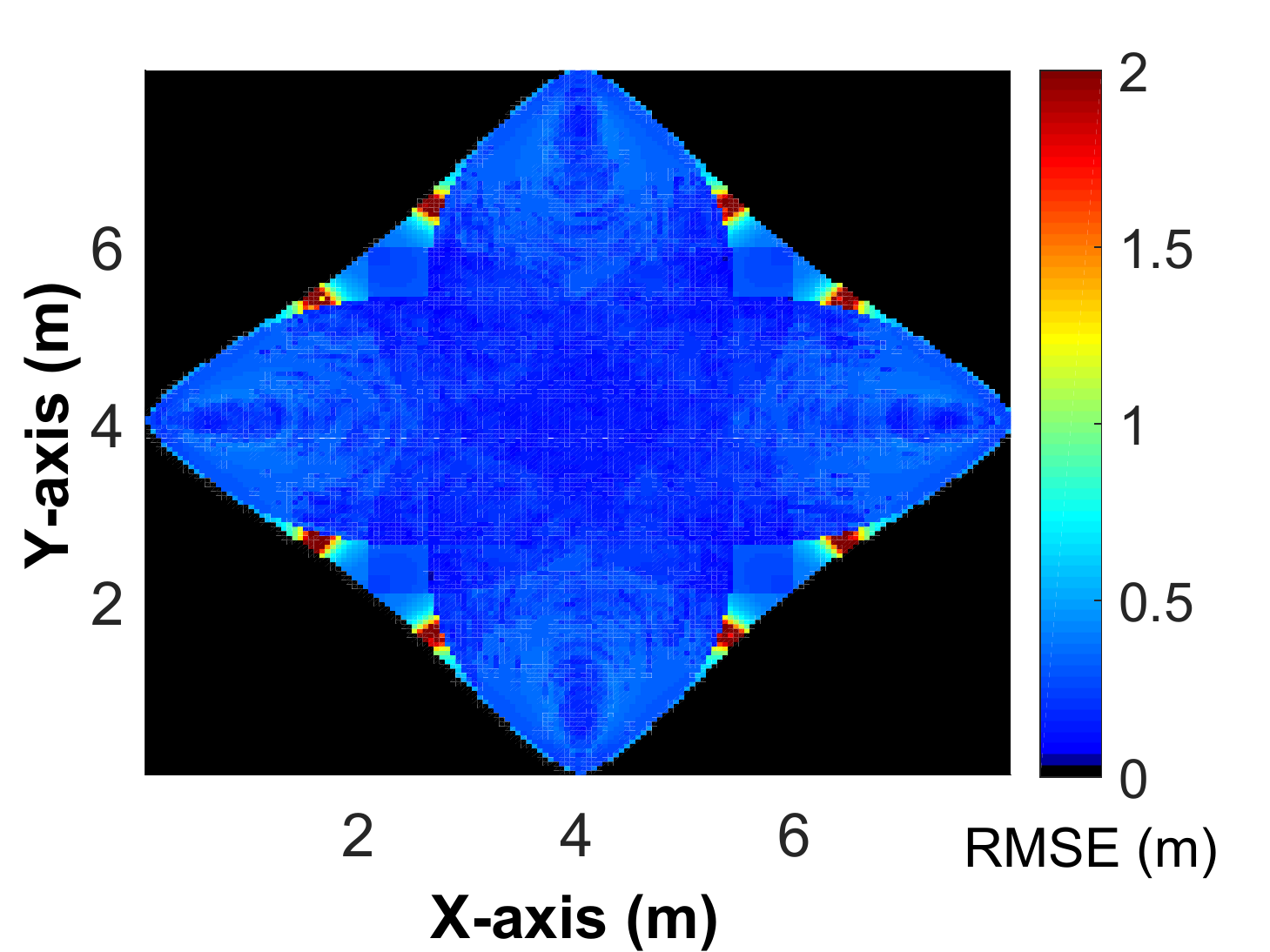}
\vspace{-5mm}
\label{fig:crlb45b}
\caption{CRLB Bistatic}
\end{subfigure}
\caption{Average MLE and CRLB RMSE for monostatic and bistatic configurations for $\theta_i=\pi/4$, $H_i=1$~meter, for $i~=~1,\dots,4$, and $\boldsymbol{\rm p}_{\rm Tx}=1000$~mW.}
\label{fig:t45mc}
\end{figure}

\begin{figure*}[t]
	\centering
	\includegraphics[width=1\linewidth]{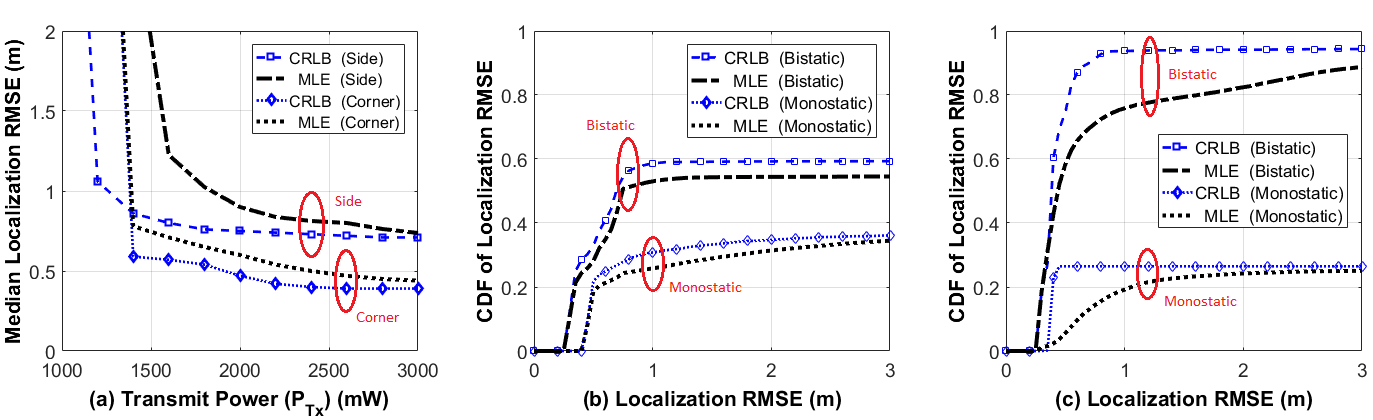}
	\caption{Deployment with $\boldsymbol{\theta_i=\pi/4}$ for $i=1,2,3,4$, (a) Median localization RMSE of MLE and CRLB for various transmit powers, (b) CDF of RMSE of MLE and CRLB with side placement ($P_{\rm Tx}~=~3000$~mW), (c) CDF of RMSE of MLE and CRLB with corner placement ($P_{\rm Tx}~=~3000$~mW).}
	\label{fig:pi4all}
	\vspace{-5mm}
\end{figure*}	

\begin{figure*}[t]
	\centering
	\includegraphics[width=1\linewidth]{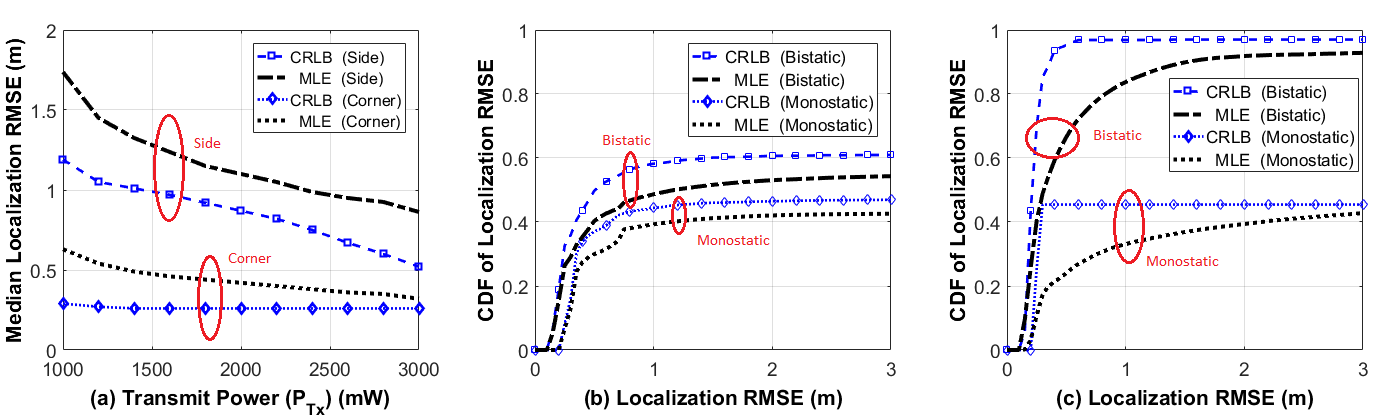}
	\caption{Deployment with $\boldsymbol{\theta_i=\pi/3}$ for $i=1,2,3,4$, (a) Median localization RMSE of MLE and CRLB for various transmit powers, (b) CDF of RMSE of MLE and CRLB with side placement ($P_{\rm Tx}~=~3000$~mW), (c) CDF of RMSE of MLE and CRLB with corner placement ($P_{\rm Tx}~=~3000$~mW).}
	\label{fig:pi3all}
\end{figure*}	

\begin{figure*}[t]
	\centering
\includegraphics[width=1\linewidth]{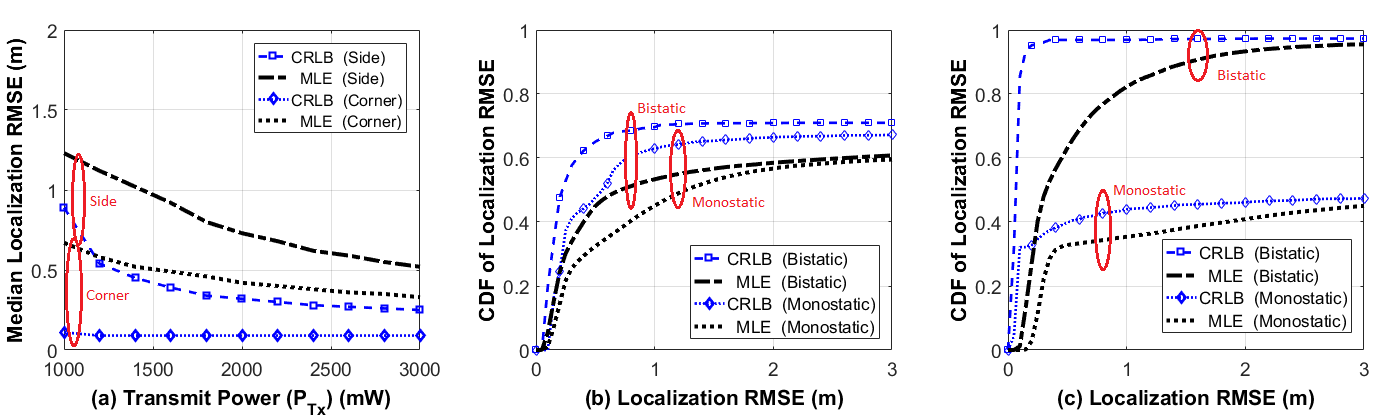}
\caption{Deployment with $\boldsymbol{\theta_i=\pi/2}$ for $i=1,2,3,4$, (a) Median localization RMSE of MLE and CRLB for various transmit powers, (b) CDF of RMSE of MLE and CRLB with side placement ($P_{\rm Tx}~=~3000$~mW), (c) CDF of RMSE of MLE and CRLB with corner placement and ($P_{\rm Tx}~=~3000$~mW).}
\label{fig:pi2all}
\vspace{-5mm}
\end{figure*}	



\subsection{Localization Accuracy}
In Fig.~\ref{fig:t45mc}, Average MLE and CRLB RMSE for monostatic and bistatic configurations with $\theta=\pi/4$, for $P_{\rm Tx}=1000$~mW at each possible tag location is given.
The localization coverage for monostatic configuration is just above $20\%$, while in bistatic configuration it is above $50\%$ as represented in Fig.~\ref{fig:CoverageP}.
Monostatic configuration has localization coverage above $50\%$ for only side placement of antennas with $\theta=\pi/2$, thus they are not represented in median localization RMSE results which they do not have.


The median localization RMSE of CRLB and MLE are compared in Fig.~\ref{fig:pi4all}(a), for elevation angle of $\theta~=~\pi/4$.
Monostatic configuration is not in the results since it does not have a coverage above $50\%$ as in Fig.~\ref{fig:CoverageP}.
Median RMSE of CRLB for side placement of antennas begin with $1.07$ meters at $P_{\rm Tx}=1200$~mW and gets as low as $0.72$ meters, while corner placement has lower median error in general from $0.61$ meters at $P_{\rm Tx}=1400$~mW to $0.43$ meters at $P_{\rm Tx}=3000$~mW.
As expected, MLE gets closer performance to the CRLB as transmit power increases.
Median RMSE of MLE for side placement of antennas begin with $1.26$ meters at $P_{\rm Tx}=1600$~mW to $0.76$ meters at $P_{\rm Tx}=3000$~mW, while corner placement does better with $0.73$ meters at $P_{\rm Tx}=1400$~mW, and $0.45$ meters at $P_{\rm Tx}=3000$~mW.

In Fig.~\ref{fig:pi4all}(b), performance of side placement, and in Fig.~\ref{fig:pi4all}(c), performance of corner placement is are shown.
In Fig.~\ref{fig:pi4all}(b), the localization probability of a tag with MLE below an error of $1$ meter for monostatic configuration with side placement and $P_{\rm Tx}=3000$~mW is $0.26$, while for bistatic configuration with same parameters it gets to $0.53$. 
The CDF values of CRLB for those are $0.31$ and $0.59$, respectively.

In Fig.~\ref{fig:pi4all}(c), the localization probability of a tag with MLE below an error of $1$ meter for monostatic configuration with corner placement and $P_{\rm Tx}=3000$~mW is $0.19$, while for bistatic configuration with same parameters it gets to $0.76$. 
The CDF values of CRLB for those are $0.26$ and $0.92$, respectively.
The side placement of antennas has better performance with monostatic MLE compared to corner placement, while bistatic performance substantially lower.

Increasing elevation angle to $\theta=\pi/3$ helps to decrease median localization RMSE and improve localization performance.
The median localization RMSE of CRLB and MLE are compared in Fig.~\ref{fig:pi3all}(a), for elevation angle of $\theta~=~\pi/3$.
As shown in Fig.~\ref{fig:CoverageP}, bistatic configuration is always above $50\%$ in localization coverage.
In Fig.~\ref{fig:pi3all}(a), median RMSE of CRLB for side placement of antennas begin with $1.21$ meters at $P_{\rm Tx}=1000$~mW and gets as low as $0.51$ meters, while corner placement has lower median error in general from $0.32$ meters to $0.3$ meters at $3000$~mW.
Similar to $\theta=\pi/4$, MLE converges to CRLB as transmit power increases.
Median RMSE of MLE for side placement of antennas begin with $1.71$ meters at $P_{\rm Tx}=1000$~mW, which reduces to $0.78$ meters at $P_{\rm Tx}=3000$~mW, while corner placement does better with $0.63$ meters and $0.34$ meters, respectively.

In Fig.~\ref{fig:pi3all}(b) CDF of localization RMSE for side placement is shown for side placement with $\theta=\pi/3$.
The localization probabilities of a tag below an error of $1$ meter for monostatic and bistatic configuration are $0.33$ and $0.84$, respectively,
while their CRLB are $0.44$ and $0.97$, respectively.

In Fig.~\ref{fig:pi3all}(c), the localization probability of a tag with MLE below an error of $1$ meter for monostatic configuration with corner placement and $P_{\rm Tx}=3000$~mW are $0.41$ and $0.51$, while their CRLB are $0.44$ and $0.58$, respectively.
Side placement of antennas increase the performance of monostatic configuration while degrading bistatic configuration performance similar to $\theta=\pi/4$.

In Fig.~\ref{fig:pi2all}(a), the median localization RMSE of CRLB and MLE are compared for elevation angle of $\theta~=~\pi/2$ with side and corner placement of antennas.
Median RMSE of CRLB for side placement of antennas begin with $0.89$ meters at $P_{\rm Tx}=1000$~mW and gets as low as $0.25$ meters, while corner placement has lower median error in general from $0.11$ meters to $0.09$ meters at $3000$~mW.
Similar to $\theta=\pi/4$ and $\theta=\pi/3$, MLE converges to CRLB as transmit power increases.
Median RMSE of MLE for side placement of antennas begin with $1.23$ meters at $P_{\rm Tx}=1000$~mW and reduce to $0.52$ meters at $P_{\rm Tx}=3000$~mW, while corner placement does better with $0.67$ meters and $0.33$ meters, respectively.

In Fig.~\ref{fig:pi2all}(b) CDF of localization RMSE for side placement is represented.
The localization probabilities of a tag below an error of $1$ meter for monostatic and bistatic configuration are $0.35$ and $0.82$, respectively.
The CRLB for those are $0.44$ and $0.97$, respectively.
In Fig.~\ref{fig:pi2all}(c), the localization probability of a tag with MLE and CRLB is shown with respect to localization RMSE. The probability of having an error below $1$ meter for monostatic configuration with corner placement and $P_{\rm Tx}=3000$~mW are $0.47$ and $0.56$, while the CDF of CRLB for those are $0.63$ and $0.70$, respectively.
Side placement of antennas increase the performance of monostatic configuration slightly while degrading bistatic configuration performance substantially.

In general, configurations with larger elevation angle results better localization coverage and lower localization RMSE.
In Fig.~\ref{fig:pi4all}(a), the median localization RMSE for $\theta=\pi/4$ has much higher values compared to $\theta=\pi/3$ in Fig.~\ref{fig:pi3all}(a) and $\theta=\pi/2$ in Fig.~\ref{fig:pi2all}(a), for example, at $P_{\rm Tx}=1000$~mW localization RMSE is not available for $\theta=\pi/4$ since its localization coverage is all below $50\%$ for either corner and side placement of antennas, while $\theta=\pi/3$ and $\theta=\pi/2$ have acceptable accuracies.
Especially $\theta=\pi/2$ has median localization RMSE of $0.5$ meters for both side and corner configuration.
At all elevation angles, corner placement of antennas has better localization coverage for bistatic configuration at $P_{\rm Tx}=3000$~mW.
Monostatic configuration does better with side placement of antennas, since in that case the coverage of antennas overlaps in larger areas.
Increasing transmit power not only increases the localization coverage, but also reduces the localization error.
As a conclusion, an elevation angle larger than $\theta=\pi/3$ is crucial for localization coverage and accuracy as well as corner placement of antennas with transmit power at $3000$~mW which is the EIRP limit in EPC Gen2 protocol of UHF RFID systems.

\section{Conclusion}
\label{sect:Conclusion}
In this paper, fundamental limits on the IoT localization accuracy of a passive UHF RFID tag is studied considering realistic propagation models for reader antennas.
Our results show that high accuracy of localization does not only depend on the transmit power, but also depends on the use of
right elevation angle and antenna placement and the use of bistatic configuration in localization system.
In our simulations it is shown that among the considered elevation angles, $\theta=\pi/2$ yields the best results for the given deployment scenario, since it maximizes the received power, results in largest localization coverage for IoT and minimizes the localization error.
We observed that bistatic localization coverage drops with the use of side placement of antennas, while it increases monostatic localization coverage.
Using bistatic configurations improves the probability of localizing the tag with higher accuracies when compared with monostatic configurations.
The best results are achieved with bistatic configuration and side placement of the antennas.

\section*{Acknowledgement}
This work was made possible by the National Science Foundation Grant AST\textendash1443999. The statements made herein are solely the responsibility of the authors.
\appendices
\section{Proof of Theorem 1}
\label{sect:Derivation}
In this appendix we will show derivation of CRLB through obtaining FIM.
	Individual elements of the FIM in \eqref{eq:fim} can be calculated using the likelihood function $\mathcal{L}( \boldsymbol{\rm \hat{p}};\mbox{$\boldsymbol{\rm x}$})$ in \eqref{eq:likelihood} as follows\cite{Estimation1993Kay}:
	\begin{align}
	\textbf{I}_{mn}=-E\left[\dfrac{\partial^2\ln \mathcal{L}( \boldsymbol{\rm \hat{p}};\mbox{$\boldsymbol{\rm x}$})}{\partial {\rm x}_m\partial {\rm x}_n}\right],
	\label{eq:fim_element}
	\end{align}
	where $\textbf{I}_{mn}$ is the $mn$-th element of the FIM for $m,n=1,2$. 
	As in \cite{Estimation1993Kay}, using \eqref{eq:likelihood} the FIM element in \eqref{eq:fim_element} can be derived as
	\begin{align}
	-E\left[\dfrac{\partial^2\ln \mathcal{L}( \boldsymbol{\rm \hat{p}};\mbox{$\boldsymbol{\rm x}$})}{\partial {\rm x}_m\partial {\rm x}_n}\right]=\frac{1}{\sigma^2}\sum_{i=1}^{N}\sum_{j=1}^{N}\bigg(\frac{\partial\hat{{P}}_{ ij}}{\partial {\rm x}_m}\times\frac{\partial\hat{{P}}_{ ij}}{\partial {\rm x}_n}\bigg).\label{eq:expectfim}
	\end{align}
	Note that \eqref{eq:Preceived2} is in logarithmic scale. 
	Derivative of each element in received power is calculated separately since it can be written as summation of different functions in logarithmic scale.
	Partial derivative of \eqref{eq:Preceived2} can be represented~as
	\begin{align}
	\frac{\partial\hat{{P}}_{ ij}}{\partial {\rm x}_m}&=\frac{\partial\big(20\log_{10}\big(\tau\mu_{\rm T}\rho_{\rm L}P_{\rm Tx}G^2_{\rm T}|h_ih_j\Gamma|^2\big)\big)}{\partial {\rm x}_m}\nonumber\\
	&+\frac{\partial\big(20\log_{10}G_{\rm R}^i\big)}{\partial {\rm x}_m}+\frac{\partial\big(20\log_{10}G_{\rm R}^j\big)}{\partial {\rm x}_m}\nonumber\\
	&+\frac{\partial\big(20\log_{10}L(d_i)\big)}{\partial {\rm x}_m}+\frac{\partial\big(20\log_{10}L(d_j)\big)}{\partial {\rm x}_m}.\label{eq:PrecDerivative}
	\end{align}
	The (unknown) location of the tag ($\boldsymbol{\rm x}$) does not affect the parameters $\tau\mu_{\rm T}\rho_{\rm L}P_{\rm Tx}G^2_{\rm T}|h_ih_j\Gamma|^2$ of received power, and hence the	resulting partial derivative of \eqref{eq:Preceived2} is then given by
	\begin{align}
	\frac{\partial\hat{{P}}_{ ij}}{\partial {\rm x}_m}=\frac{20}{\ln10}\bigg(\frac{\partial G_{\rm R}^i}{\partial {\rm x}_m}+\frac{\partial G_{\rm R}^j}{\partial {\rm x}_m}+\frac{\partial L(d_i)}{\partial {\rm x}_m}+\frac{\partial L(d_j)}{\partial {\rm x}_m}\bigg).\label{eq:partialdx}
	\end{align}
\section{Example Derivation for CRLB}
\label{sect:DerivationExample}
In this appendix we will derive the CRLB for parameters $\theta_i=\pi/4$ and $\phi_i=\pi/2$ for $i=1,2,3,4$.
  The gain function in \eqref{eq:patchgaincartesian} for those particular values of $\theta_i$ and $\phi_i$ becomes
\begin{align}
G^i_R&=3.136\times\sin^2\bigg(\frac{\sqrt{2}\pi(l_i+z_i-z_0)}{4 d_i}\bigg)\nonumber\\
&\times\cos^2\bigg(\frac{\sqrt{2}\pi(l_i-z_i+z_0)}{4 d_i}\frac{(y_i-y_0)}{l_i}\bigg).\label{eq:G_Rexample}
\end{align}
First derivative of \eqref{eq:G_Rexample} with respect to ${\rm x}_m$, for $m=1,2$, is
\begin{align}
\frac{\partial G^i_R}{\partial {\rm x}_m}=\dfrac{3.136}{4}\times \frac{\partial B}{\partial {\rm x}_m} \sin A \sin (2B) \times \frac{\partial A}{\partial {\rm x}_m}\sin (2A)\cos B\label{eq:Gderexample}
\end{align}
where
\begin{align}
A=&\frac{\sqrt{2}\pi(l_i+z_i-z_0)}{4 d_i},\nonumber\\
B=&\frac{\sqrt{2}\pi(l_i-z_i+z_0)}{4 d_i}\frac{(y_i-y_0)}{l_i}.\nonumber
\end{align}
Then for ${\rm x}_1=x$ in \eqref{eq:Gderexample}, $\frac{\partial A}{\partial x}$ and $\frac{\partial B}{\partial x}$ can be solved as
\begin{align}
\frac{\partial A}{\partial x}&=\frac{\pi}{2\sqrt{2}}\bigg(\frac{x-x_i}{ l_i d_i}-\frac{(x-x_i)(l_i+z-z_i)}{d_i^3}\bigg),\nonumber\\
\frac{\partial B}{\partial x}&=\frac{\pi(x-x_i)(y-y_i)}{2\sqrt{2}l_i d_i}\bigg(\frac{1}{l_i}-\frac{(l_i-z+z_i)(l_i^2+d_i^2)}{l_i^2 d_i^2}\bigg).\nonumber
\end{align}
The same solution for ${\rm x}_2=y$ is given in
\begin{align}
\frac{\partial A}{\partial y}&=\frac{\pi}{2\sqrt{2}}\bigg(\frac{y-y_i}{l_i d_i}-\frac{(y-y_i)(l_i+z-z_i)}{d_i^3}\bigg),\nonumber\\
\frac{\partial B}{\partial y}&=\frac{\pi(y-y_i)(l_i-z+z_i)}{2\sqrt{2}l_i d_i}\nonumber\\&\times\bigg(\frac{1}{l_i(l_i-z+z_i)}-\frac{1}{l_i^2}-\frac{1}{d_i^2}-\frac{1}{(y-y_i)^2}\bigg).\nonumber
\end{align}
The path loss function $L(d_i)$ does not change with $\theta$ and $\phi$, and it only depends on the distance between the reader antenna and the tag. 
Then, the derivative of $L(d_i)$ with respect to $x$ and $y$ is as follows
\begin{align}
\frac{\partial L(d_i)}{\partial x}=\frac{\lambda^2}{(4\pi)^2}\bigg(\frac{x-x_i}{d_i^3}\bigg),\nonumber\\
\frac{\partial L(d_i)}{\partial y}=\frac{\lambda^2}{(4\pi)^2}\bigg(\frac{y-y_i}{d_i^3}\bigg).\nonumber
\end{align} 
Based on these derivations, using \eqref{eq:CRLB}--\eqref{eq:partialdx}, the CRLB for any location can be calculated with known set $x_i$ and $y_i$ for $i = 1,...,N$ with given parameters $\theta_i=\pi/4$ and $\phi_i=\pi/2$.
In Fig.~\ref{fig:t45mc}(c) and Fig.~\ref{fig:t45mc}(d), CRLB for monostatic and bistatic configurations respectively are calculated for any possible location of tag.

\bibliographystyle{IEEEtran}

\end{document}